# Atmosphere Extinction at the ORM on La Palma: A 20 yr Statistical Database Gathered at the Carlsberg Meridian Telescope

Alejandro García-Gil

Instituto de Astrofísica de Canarias, c/ Vía Láctea s/n, E-38205, La Laguna, Tenerife; and
Agencia Estatal de Meteorología (AEMET), Spain; agg@iac.es

Casiana Muñoz-Tuñón

Instituto de Astrofísica de Canarias, c/ Vía Láctea s/n, E-38205, La Laguna, Tenerife; and Departamento de Astrofisica,
Universidad de La Laguna, E-38205, La Laguna, Tenerife, Spain; cmt@iac.es

and

Antonia M. Varela

Instituto de Astrofísica de Canarias, c/ Vía Láctea s/n, E-38205, La Laguna, Tenerife; and Departamento de Astrofisica,
Universidad de La Laguna, E-38205, La Laguna, Tenerife, Spain; avp@iac.es

**ABSTRACT.** The Observatorio del Roque de los Muchachos (ORM), in the Canary Islands (Spain), was one of the candidates to host the future European Extremely Large Telescope (E-ELT) and is the site of the Gran Telescopio Canarias (GTC), the largest optical infrared facility to date. Sky transparency is a key parameter as it defines the quality of the photometry to be acquired in the astronomical observations. We present a study of the atmosphere extinction at the ORM, carried out after analysis of a database spanning more than 20 yr,[1] to our knowledge, the longest and most complete and homogeneous in situ database available for any observatory. It is based on photometric measurements in the $V$ band and $r'$ band (transformed to the $V$-band extinction coefficient $k_V$) using the Carlsberg Meridian Telescope (CMT). Clear seasonal variations that repeat yearly are observed. The median value of $k_V$ is 0.13 mag airmass$^{-1}$; the mean value has a maximum in the summer months (June–September), corresponding to the season with maximum frequency of nights affected by dust or cirrus (∼29% in summer, but only ∼13% during the rest of the year). Two volcanic eruptions took place during the database baseline, which has enabled the study of the impact of volcanoes on the global atmosphere extinction. For the 5 yr of available information, we have estimated the average monthly weather downtime from the CMT data log, obtaining a result (20.7%) in reasonable agreement with earlier studies. The main conclusion of our study is that there is no significant evidence from the CMT data for any secular changes in $k_V$ over the 20 yr database baseline.

## 1. INTRODUCTION

The Canarian Observatories, at La Palma (Observatorio del Roque de Los Muchachos, ORM) and Tenerife (Observatorio del Teide, OT), Canary Islands, Spain, host telescopes and instruments of over 60 scientific institutions from 19 different countries which have carried out permanent astronomical observation for over 40 years. The summits of Tenerife, where the OT is sited, have been a reference site for atmospheric studies since the end of the nineteenth century. The Canarian Observatories host first-class telescopes such as the 10 m Gran Telescopio Canarias (GTC) at ORM and the Solar VTT (Vacuum Tower Telescope) at OT. Moreover, the ORM has been one of the candidate sites for hosting the European Extremely Large Telescope (E-ELT) (see Muñoz-Tuñón et al. 2007; Vernin et al. 2008 and references therein) and was also finalist for hosting the Advanced Solar Telescope (ATST). The European Solar Telescope (EST), currently in advanced design status, is to be erected either at ORM or at OT.

For nighttime, the ORM has been monitored and characterized over several decades (for a review, see Muñoz-Tuñón 2002 and Muñoz-Tuñón et al. 2007). Over 20 yr worth of data have been published, confirming the world class astronomical characteristics with a stable, well known, and predictable atmosphere. Site campaigns with specific instruments are routinely performed using, for example, classical Differential Image Motion Monitors (DIMM) to measure seeing (Vernin & Muñoz-Tuñón 1995; Muñoz-Tuñón et al. 1997) and Automatic Weather Stations (Varela & Muñoz-Tuñón 2009). The transparency of the air, sky background, cloudiness, etc., are also monitored.

---

[1] At http://www.ast.cam.ac.uk/~dwe/SRF/camc.html.

An updated Web site hosted by the Instituto de Astrofisica de Canarias includes summaries, compendia, databases, and online data.[2]

The main goal of the present article is analysis of the atmosphere extinction at ORM. Atmosphere extinction is the astronomical parameter that evaluates sky transparency. Sources of sky transparency degradation are clouds (water vapor) and aerosols (dust particles included). Extinction values and their stability throughout the night are essential for determining the accuracy of astronomical measurements. The nights with low and constant extinction are classified as photometric, and extinction is considered among those parameters most relevant for characterizing an observing site.

The trade wind scenario and the cold oceanic stream, in combination with the local orography, play an important role in the retention of low cloud layers well below the summits to the windward (north) of the Canary Islands above which the air is dry and stable (the cloud layers also trap a great deal of light pollution and aerosols from low troposphere).

The semipermanent trade winds in the North Atlantic produce a stable subsiding maritime air mass at low altitude, giving rise to a low inversion layer, with its mean base at 00 GMT at 800 m in July and August and 1600 m during the rest of the year (Dorta 1996), and producing a stable laminar air flux at the altitude of the observatories, which are situated well above this altitude. The trade winds, however, are not permanent; their behavior is seasonal and they occur with greater frequency in the spring and summer (from April to September) than in the autumn and winter. The summer is also the season when winds carrying dust from the Sahara Desert to the Canary Islands occur more frequently, thus increasing the atmosphere extinction when they reach the altitude of the observatories (∼2400 m). During the rest of the year, occasional Saharan dust intrusions are produced by the so-called anticyclonic glooms (Varela et al. 2008); these are less frequent and occur mostly in the lower troposphere, below the thermal inversion layer at about 1500 m. For this reason, in the analysis of this article, we consider two separate periods, the summer months (June, July, August, and September) and the rest of the year.

ORM is located along the northern ridge of the Taburiente caldera, on the northwestern side of La Palma. At 2396 m above sea level, its geographical coordinates are 17°52′34″ west and 28°45′34″ north. The ORM has been thoroughly monitored, so we have a comprehensive knowledge of the observing conditions that allows us to establish whether there are trends due to such factors as climate variations. The Carlsberg Meridian Telescope (CMT, formerly the Carlsberg Automatic Meridian Circle), a 17.8 cm diameter telescope, started operations in 1984 May and has since then provided continuous automated measurements of atmosphere extinction. We have used this database to compile long-term statistics of the atmosphere extinction. Moreover, when available, the telescope observing logs have been used to derive the percentage of weather downtime due to adverse meteorological conditions.

It is also important to note that the possible influences of climate change on the atmospheric parameters of interest for astronomical sites are currently being debated. In particular we question in this article whether there have been variations in the atmosphere extinction over the years and also whether the incidence of dust episodes at the altitude of the ORM has evolved over the last 20 yr.

The paper is structured as follows. The data are presented in § 2. A statistical analysis of the atmosphere extinction, monthly and seasonal trends, and global values is presented in § 3. A special mention is made of the issue of the CMT detection of volcanic ashes from faraway volcanoes such as Mount Pinatubo in the Philippines and El Chichón in Mexico. The results from the analysis of the data set affected by the volcanoes and their effect on the extinction are also included in § 3. A specific study of the frequency of nights affected by dust or cirrus is presented in § 4. In § 5 an analysis of the weather downtime is carried out. Finally, in § 6, the conclusions of the work and a discussion comparing our results with those of other studies are presented.

## 2. DESCRIPTION OF THE DATA AND DATABASE

Atmosphere extinction, relevant to ground-based astronomy in optical and near-infrared wavelengths, is associated with the absorption/scattering of incoming photons from astronomical sources by the Earth's atmosphere: Rayleigh scattering by air atoms and molecules (around 600 nm); Mie scattering by aerosols (Tüg et al. 1977); and ozone absorption (below 320 nm). Discrete telluric absorption lines arise from molecular oxygen, ozone, and water molecules. In astronomy, the atmosphere extinction is measured by the extinction coefficient $k$, which is wavelength, altitude, and time dependent, and can be determined by making multiple observations of stars at different air masses (a function of the angle above the horizon) along a given night.

The total extinction coefficient $k(\lambda)$ at zenith in clear nights adopted by Hayes & Latham (1975) is given by

$$k_\lambda = k_{\text{Ray}}(\lambda, h) + k_{\text{aer}}(\lambda, h) + k_{\text{oz}}(\lambda). \quad (1)$$

$k_{\text{Ray}}(\lambda, h)$ is the Rayleigh vertical scattering and depends on wavelength ($\lambda$) and altitude ($h = 2369$ m at the ORM); $k_{\text{oz}}(\lambda)$ is the molecular absorption by ozone which depends on the absorption coefficient (Gast 1980) and on the total ozone over ORM (from curves appropriate to the latitude of La Palma, Allen 1963); and $k_{\text{aer}}(\lambda, h)$ is the aerosol scattering.

Following Jones (1984), dust scattering at ORM does not strongly depend on wavelength. Then, the dust correction term to the extinction curve can be determined comparing the observational extinction coefficient $K(V)$ provided by the CMT and the theoretical extinction given by $(k_{\text{Ray}}[V] + k_{\text{oz}}[V])$. For more details, see Gutiérrez-Moreno (1982) and King (1985).

---

[2] At http://www.iac.es/site-testing/.

The CMT started measuring the atmosphere extinction in the V band ($k_V$) on 1984 May 13, finishing the measurements on 1998 May 28, with an observational uncertainty of $\pm 0.005$ mag airmass$^{-1}$. On 1999 March 26, the filter was changed to $r'$ (with an effective wavelength of 625 nm), and measurements are now obtained with an observational uncertainty of $\pm 0.0005$ mag airmass$^{-1}$; $k_{r'}$ can be transformed into $k_V$ by adding a factor of $\pm 0.029$ mag airmass$^{-1}$ (King 1985). In that manner, both the $k_V$ data and $k_{r'}$ transformed into $k_V$ data can be compared together.

There are two periods in the analyzed database when the atmospheric transparency above the ORM decreases due to the increase in aerosol density associated with high volcanic activity elsewhere: the eruption of El Chichón in Mexico on 1982 March 29 and the eruption of Mount Pinatubo in the Philippines on 1991 June 12. The two periods of high atmosphere extinction associated with these volcanic eruptions have not been taken into consideration in the trends analysis (that is, from 1984 May to 1988 February and from 1991 June to 1993 December). There is also a period without data due to the changeover from the V to the $r'$ filter and to resetting and calibration work at the telescope.

A specific study of the effect of the volcanoes at the ORM altitude is included in § 3.2.

An extensive statistical analysis of the atmosphere extinction in the V band has been carried out. For the $k_V$ data, a binning of $\pm 0.01$ mag airmass$^{-1}$ has been used in order to calculate the median for grouped data ($\tilde{K}_V$), defined as in Spiegel (1988)

$$\tilde{K}_V = L_1 + \left(\frac{\frac{N}{2} - (\sum f)_1}{f_{\text{median}}}\right)C. \qquad (2)$$

Here $L_1$ is the inferior border of the median class, $N$ is the number of data, $(\sum f)_1$ is the sum of frequencies of classes inferior to that of the median, $f_{\text{median}}$ is the frequency of the median class, and $C$ is the range of the interval of the median class. The mode for grouped data ($\hat{K}_V$), defined as in Spiegel (1988) is

$$\hat{K}_V = L_1 + \left(\frac{\Delta_1}{\Delta_1 + \Delta_2}\right)C, \qquad (3)$$

where $L_1$ is the inferior border of the median class, $\Delta_1$ is the excess of modal frequence over the immediate inferior class, $\Delta_2$ is the excess of modal frequence over the immediate superior class, and $C$ is the range of the interval of the median class.

## 3. DATA ANALYSIS

The mean and median $k_V$ values for the months included in the database have been computed. The median monthly values are summarized in Table 1 and shown in Figure 1 with the typical error of the median between parenthesis. To calculate this error, we have assumed a normal distribution, using the formula (Spiegel 1988)

$$\Delta \tilde{K}_V = \frac{1.2533s}{\sqrt{(N)}}, \qquad (4)$$

where $s$ is the typical deviation of the data and $N$ is the number of nights. The statistics have been performed for months with more than 5 nights of observations so as exclude months with too small a sample (12 months are discarded for this reason, 4 of which have no data). Regarding the monthly behavior, from results compiled in Table 1 and shown in Figure 1, the small variation in the monthly $k_V$ median during the year is clear. The mean monthly median value is 0.13 mag airmass$^{-1}$, the monthly median being between 0.121 and 0.133 mag airmass$^{-1}$, except for August, when the median value is 0.142 mag airmass$^{-1}$. There are 6 months with a median above or equal to 0.2 mag airmass$^{-1}$: 4 of these are in month 8 (August of years 1988, 1990, 2000, and 2008) and 2 in month 7 (July of years 1994 and 2009). All of them are in the summer and are produced by a persistent Saharan dust plume at the midheight (1.5–5 km) of the troposphere.

### 3.1. Trends and Global Values

As mentioned before, the median $k_V$ value for each month is fairly constant throughout the year (see the line in Fig. 1). The mean value, however, is expected to increase in summer due to the higher frequency of Saharan dust episodes at the altitude of the observatories during this season. To explore this *seasonal behavior*, as discussed in § 1, the study has been divided in two blocks: summer (June–September), and the rest of the year.

The cumulative frequency of $k_V$ is shown in Figure 2 for the entire period under study. The similarity is evident between the statistics in the summer and the rest of the year until ~55%, the median value thus being included. There are differences for higher extinction. The median $k_V$ is 0.131 mag airmass$^{-1}$ in summer and 0.129 during the rest of the year, values that are identical to within errors. The modal $k_V$ is 0.120 mag airmass$^{-1}$ during summer and 0.121 during the rest of the year, also identical to within errors. However, the mean $k_V$ is 0.183 mag airmass$^{-1}$ in summer and 0.144 for the rest of the year. The summary is shown in Table 2.

The possible influence of the filter change (V to $r'$) and the observing uncertainty change ($\pm 0.005$ to $\pm 0.0005$ mag airmass$^{-1}$) in the measurements is analyzed, although the prescribed conversion factor and an adequate binning and statistics for grouped data have been applied. We have observed that the

TABLE 1
Median $k_V$ During the Period Not Affected by Mount Pinatubo and El Chichón Eruptions

| Year | 1 | 2 | 3 | 4 | 5 | 6 | 7 | 8 | 9 | 10 | 11 | 12 |
|---|---|---|---|---|---|---|---|---|---|---|---|---|
| 1988 | | | | | 0.120(0.014) | 0.131(0.020) | 0.145(0.030) | 0.242(0.050) | 0.143(0.031) | 0.112(0.007) | | 0.120(0.040) |
| 1989 | 0.109(0.010) | 0.121(0.019) | 0.133(0.027) | 0.135(0.007) | 0.140(0.010) | 0.123(0.021) | 0.175(0.053) | 0.195(0.035) | 0.125(0.011) | 0.123(0.025) | | |
| 1990 | 0.112(0.003) | 0.125(0.013) | 0.110(0.031) | 0.111(0.024) | 0.111(0.005) | 0.104(0.007) | 0.140(0.028) | 0.245(0.042) | 0.115(0.049) | 0.119(0.012) | 0.112(0.013) | |
| 1991 | 0.115(0.009) | 0.123(0.007) | 0.116(0.006) | 0.141(0.009) | 0.135(0.030) | | | | | | | |
| 1992 | | | | | | | | | | | | |
| 1993 | | | | | | | | | | | | |
| 1994 | 0.137(0.012) | 0.143(0.009) | 0.145(0.009) | 0.167(0.030) | 0.146(0.005) | 0.136(0.031) | 0.317(0.041) | 0.140(0.017) | 0.124(0.032) | 0.121(0.007) | 0.129(0.008) | 0.125(0.012) |
| 1995 | 0.127(0.009) | 0.138(0.034) | 0.155(0.044) | | 0.138(0.006) | 0.134(0.009) | 0.137(0.024) | 0.155(0.019) | 0.124(0.020) | 0.140(0.016) | 0.127(0.020) | 0.115(0.016) |
| 1996 | 0.138(0.010) | 0.127(0.005) | 0.122(0.004) | 0.148(0.013) | 0.150(0.028) | 0.132(0.003) | 0.139(0.036) | 0.129(0.027) | 0.138(0.006) | 0.130(0.010) | 0.140(0.025) | 0.117(0.009) |
| 1997 | 0.114(0.004) | 0.168(0.021) | | 0.122(0.004) | 0.129(0.005) | 0.114(0.004) | 0.123(0.010) | 0.106(0.009) | | 0.115(0.010) | 0.119(0.014) | 0.113(0.005) |
| 1998 | 0.123(0.006) | | 0.138(0.034) | 0.160(0.011) | 0.152(0.004) | 0.159(0.039) | 0.128(0.039) | 0.146(0.031) | 0.128(0.002) | | | |
| 1999 | | 0.130(0.009) | 0.125(0.007) | 0.131(0.019) | 0.124(0.014) | 0.126(0.057) | 0.121(0.033) | 0.206(0.032) | 0.132(0.013) | 0.134(0.008) | 0.156(0.014) | 0.128(0.004) |
| 2000 | 0.120(0.018) | 0.137(0.018) | 0.188(0.050) | 0.125(0.005) | 0.130(0.006) | 0.126(0.012) | 0.121(0.031) | 0.137(0.028) | | | | |
| 2001 | 0.118(0.002) | 0.137(0.018) | 0.129(0.004) | 0.127(0.003) | 0.129(0.010) | 0.133(0.013) | 0.138(0.033) | 0.139(0.023) | 0.129(0.020) | 0.160(0.036) | 0.136(0.004) | 0.145(0.014) |
| 2002 | 0.114(0.008) | 0.112(0.020) | 0.121(0.003) | 0.148(0.072) | 0.148(0.010) | 0.132(0.017) | 0.109(0.039) | 0.121(0.014) | 0.127(0.021) | 0.164(0.018) | 0.149(0.004) | 0.152(0.018) |
| 2003 | 0.136(0.031) | 0.126(0.014) | 0.143(0.013) | 0.123(0.018) | 0.128(0.017) | 0.126(0.012) | 0.150(0.051) | 0.132(0.021) | 0.153(0.026) | 0.124(0.020) | 0.112(0.002) | 0.115(0.014) |
| 2004 | 0.112(0.004) | 0.138(0.014) | 0.135(0.025) | 0.135(0.006) | 0.137(0.006) | 0.132(0.027) | 0.125(0.029) | 0.129(0.031) | 0.164(0.020) | 0.145(0.018) | 0.140(0.020) | 0.121(0.002) |
| 2005 | 0.123(0.003) | | 0.117(0.010) | 0.121(0.013) | 0.127(0.021) | 0.118(0.011) | 0.117(0.016) | 0.125(0.038) | 0.143(0.037) | 0.137(0.006) | 0.145(0.012) | 0.138(0.008) |
| 2006 | 0.126(0.001) | 0.124(0.004) | 0.121(0.025) | 0.133(0.010) | 0.125(0.007) | 0.114(0.005) | 0.129(0.030) | 0.128(0.037) | 0.126(0.023) | 0.149(0.014) | 0.132(0.011) | 0.135(0.008) |
| 2007 | 0.141(0.015) | 0.116(0.004) | 0.120(0.005) | | 0.133(0.010) | 0.113(0.004) | 0.128(0.053) | 0.118(0.012) | 0.112(0.012) | 0.119(0.005) | | |
| 2008 | | | | 0.128(0.035) | 0.138(0.008) | 0.120(0.025) | 0.144(0.020) | 0.279(0.026) | 0.111(0.015) | 0.104(0.005) | 0.121(0.010) | 0.115(0.004) |
| 2009 | 0.112(0.004) | 0.125(0.014) | 0.125(0.005) | 0.126(0.006) | 0.119(0.003) | 0.116(0.006) | 0.247(0.033) | 0.137(0.024) | 0.134(0.017) | | | |

Note.—Months shown as 1, January; 2, February; etc. Typical deviation of the median given in parentheses.

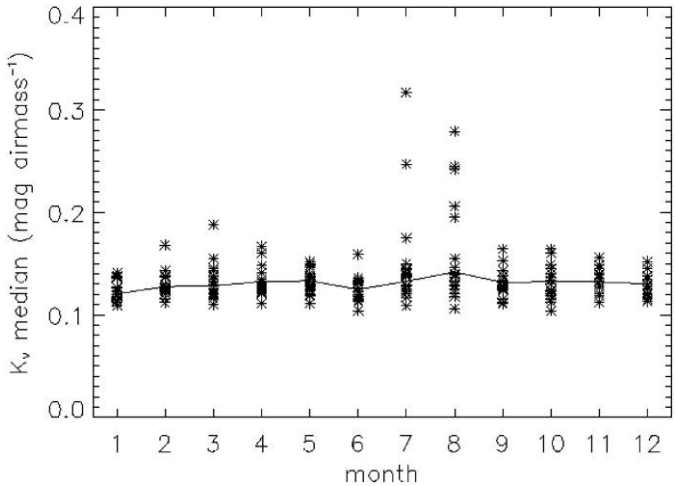

FIG. 1.—Monthly $k_V$ medians (January is month 1, February month 2, etc.) for the ~20 yr under study. The *solid line* joins the median $k_V$ for each month.

TABLE 2
SUMMARY OF $k_V$ STATISTICS (IN mag airmass$^{-1}$)

| Parameter | Summer | Rest of the Year | All Data |
|---|---|---|---|
| Ndata ...... | 1685 | 2351 | 4036 |
| Median ..... | 0.131 | 0.129 | 0.130 |
| Mode ...... | 0.120 | 0.121 | 0.121 |
| Mean ....... | 0.183 | 0.144 | 0.161 |
| Std ......... | 0.119 | 0.061 | 0.092 |

cumulative frequency of $k_V$ using only extinction obtained with a $V$ filter is very similar to that obtained using only extinction obtained with an $r'$ filter. Using the $V$-band data only, we obtain a median value of 0.131 mag airmass$^{-1}$, and 0.130 using the $r'$-band data transformed into $V$-band extinction, values which are identical to within errors. This result confirms that there is no need to apply the second correction that Guerrero et al. (1998) had suggested.

In this section we perform an analysis to detect possible trends during the years of measurements. Regarding the seasonal behavior, Figure 3 shows the monthly median and the monthly mean of $k_V$ for all the years studied. An increase in the mean for the summer is observed due to the higher frequency of Saharan dust episodes (see also Fig. 2). Figure 3 shows that the seasonal variation in the extinction repeats yearly.

To evaluate possible secular changes over the 20 yr, the yearly $k_V$ median with errors has been computed and is shown in Figure 4. Note that only years with more than 100 nights of observation are considered, so 1998 has not been included. We have made a least-squares fit and calculated the regression line slope. The standard error of the slope is higher than the slope itself, so it seems reasonable to accept the hypothesis of constant $k_V$.

As a result, it can be concluded that the median yearly extinction has not increased from 1988 to 2009. In fact, the values obtained from the line fit are approximately 0.131 and 0.128 mag airmass$^{-1}$ for 1988 and 2009, respectively. A slightly lower extinction is observed in recent years (from 2004), although this behavior is not statistically significant. From the observations, it is then clear that the net extinction at the altitude of the ORM has not changed over the 20 yr database baseline.

The complete statistical analysis using the entire database is presented in Table 2 and shown in Figure 5. The figure displays the relative frequency of $k_V$ for each bin range (width = 0.01 mag airmass$^{-1}$) for the entire period of the study. In summary, using extinction data provided by the CMT for 4036 nights, we obtain a $k_V$ median of 0.130, a mode of 0.121, a mean of $0.161 \pm 0.002$ mag airmass$^{-1}$ and a standard deviation, $s$, of 0.092 mag airmass$^{-1}$.

### 3.2. Extinction Produced by El Chichón and Mount Pinatubo Volcanoes

The impact of the two volcanoes has been quantified by analyzing the data set for the period affected. The eruption of El Chichón in Mexico in 1982 March and that of Mount Pinatubo in the Philippines in 1991 June are clearly tracked in the data on periods from 1984 May to 1988 February and from 1991 June to 1993 December. Note that El Chinchón erupted 2 yr before CMT started operating. In Figure 6 we represent the monthly mean coefficient extinction of the two periods associated to high volcanic activity elsewhere. The figure also displays, for reference, the median value of the extinction at the Observatory (0.13 mag airmass$^{-1}$) (see § 3.1)

El Chichón ashes were detected at the ORM in 1984 May at the beginning of CMT operation, almost 2 yr after the main eruption. The extinction measured is 0.2 mag airmass$^{-1}$, 0.07 mag above the median value at ORM. From then, on the tendency is to decrease although with several maxima (of the order of 0.2 mag airmass$^{-1}$). The impact of El Chichón disappears in 1988, 6 yr after the eruption. The extinction peaks in 1984 August (0.4 mag airmass$^{-1}$) and in 1985 August (0.3 mag airmass$^{-1}$) are contaminated by Saharan dust intrusions. They are coincident with large values of the total suspended particles (TSP) at the ORM obtained with the BSC-DREAM (Barcelona Supercomputing Center/Dust Regional Atmospheric Model) that delivers dust forecasts for North Africa, the Middle East, and Europe (Cuevas & Baldasano, 2009).

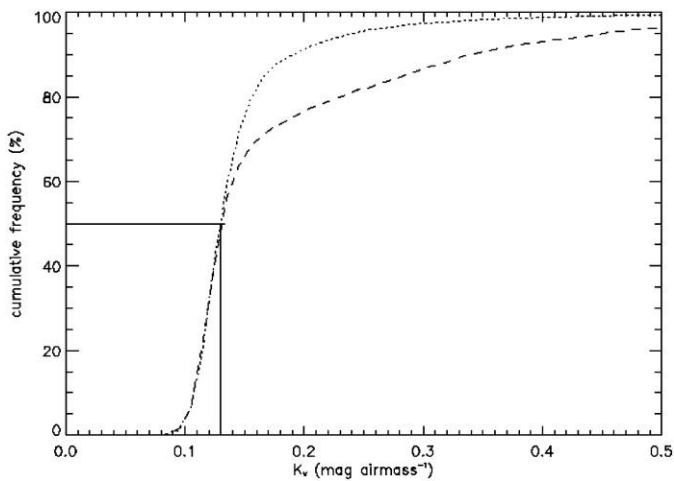

FIG. 2.—Seasonal trend: Cumulative frequency of $k_V$ for the summer months (June–September: *dashed line*) and the rest of the year (*dotted line*). Note that the median value is 0.13 both in the summer months and during the rest of the year.

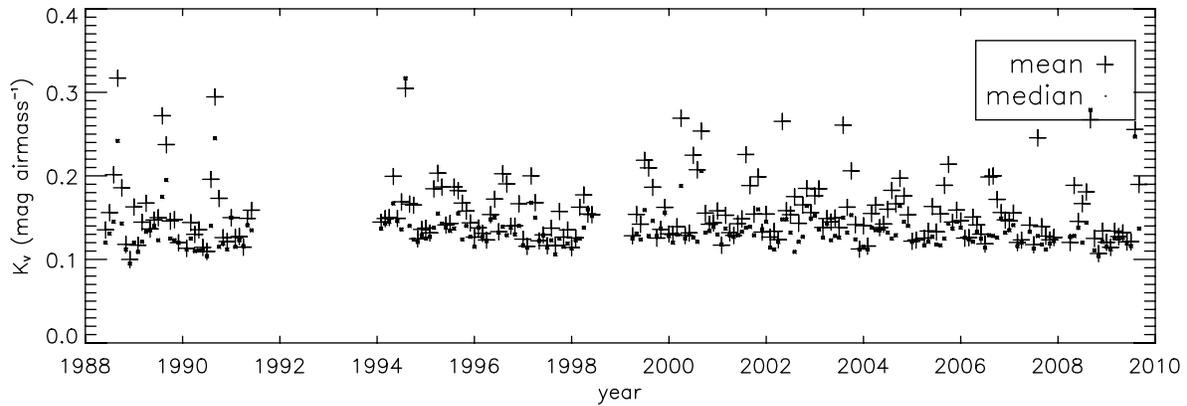

FIG. 3.—Monthly mean and median $k_V$.

The Pinatubo eruption was fully tracked, and it is clearly detected by the CMT as a sudden increase of $k_V$ (0.25 mag airmass$^{-1}$) in 1991 July, almost simultaneously with the eruption itself. The value of the extinction (0.12 mag airmass$^{-1}$ above the median value at the ORM) remained almost constant for about 1.5 yr after the eruption. Then, about 2 yr later, in 1993, the effect of the volcanic ashes started to decrease monotonically until they fully disappear in the middle of 1994, that is, 3 yr after the eruption. The values provided by the CMT during the periods affected by the two volcanoes are given in Table 3. The results provided in this section show that the CMT can be a powerful facility for quantifying the impact of volcanoes. The long database provided by the telescope allows us to identify "peculiar" features that can be due to phenomena such as the intrusion of volcanic ashes at the ORM altitude, 2392 m.

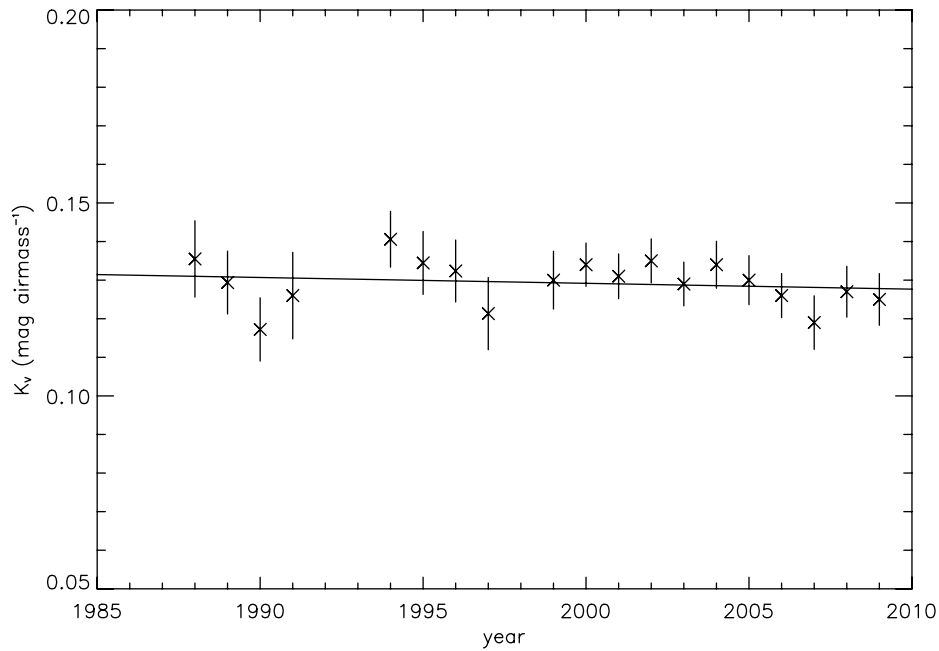

FIG. 4.—Yearly median and error bars of $k_V$ and least-squares fit with error (*dashed lines*) overplotted. Note that the fit is consistent with a constant value over the period shown of 0.13 mag airmass$^{-1}$.

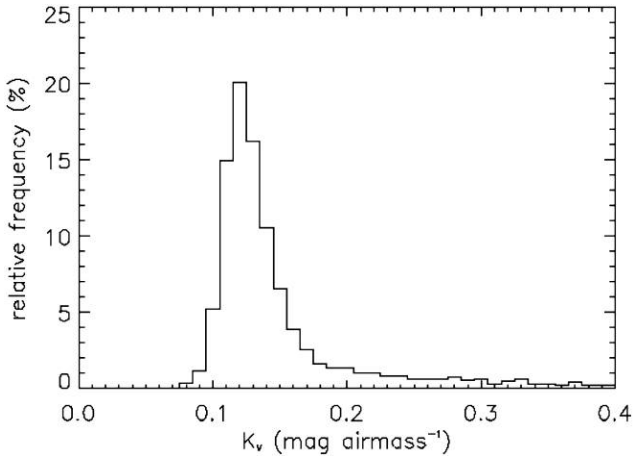

FIG. 5.—Relative frequency of $k_V$ for each range of the bin. Statistics compiled in Table 4.

## 4. STUDY OF THE NIGHTS AFFECTED BY DUST (OR CIRRUS)

In this section we determine the typical extinction values (and the percentage) of nights affected by dust and/or cirrus (as both are indistinguishable with our observations). For this aim, we assume that nonsummer nights, most of them with very low atmosphere extinction in $V$, are free of dust. The base for this assumption, common for other physical processes and biological mechanisms, is that many measurements show skewed distributions, particularly when mean values are low, variances are large, and values are not negative. Such skewed distributions often closely fit a log-normal distribution (Limpert et al. 2001). An example is provided by Matthias et al. (2004), showing that the frequency of the aerosol optical depth in the planetary boundary layer (valid for La Palma Observatory) follows a log-normal distribution at most sites. Thus we have matched the relative frequency ($f$) of $k_V$ observed ($x$) in nights free of dust to a log-normal curve, with the probability density function

$$f(x) = \frac{1}{(x-x_0)\sigma\sqrt{2\pi}} \exp\left[-\frac{(\ln(x-x_0)-\mu)^2}{2\sigma^2}\right]. \quad (5)$$

We have taken as $x_0$ the minimum $k_V$ observed (0.08 mag airmass$^{-1}$) minus the size of the bin (0.01 mag airmass$^{-1}$), i.e., $x_0 = 0.07$ mag airmass$^{-1}$.

When trying to match the fit, an excess is observed beyond ~0.185 mag airmass$^{-1}$ due to the occasional presence of dust, cirrus, etc. To have the best fit, we have considered extinction values $k_V \leq 0.205$ mag airmass$^{-1}$ only. The number of nights used in the analysis is 2170 out of the total of 2351 nights in the period. Figure 7 shows a histogram of such a comparison for nonsummer nights. A very good match is observed for $k_V < 0.185$ mag airmass$^{-1}$, and from 0.185 the differences start to increase.

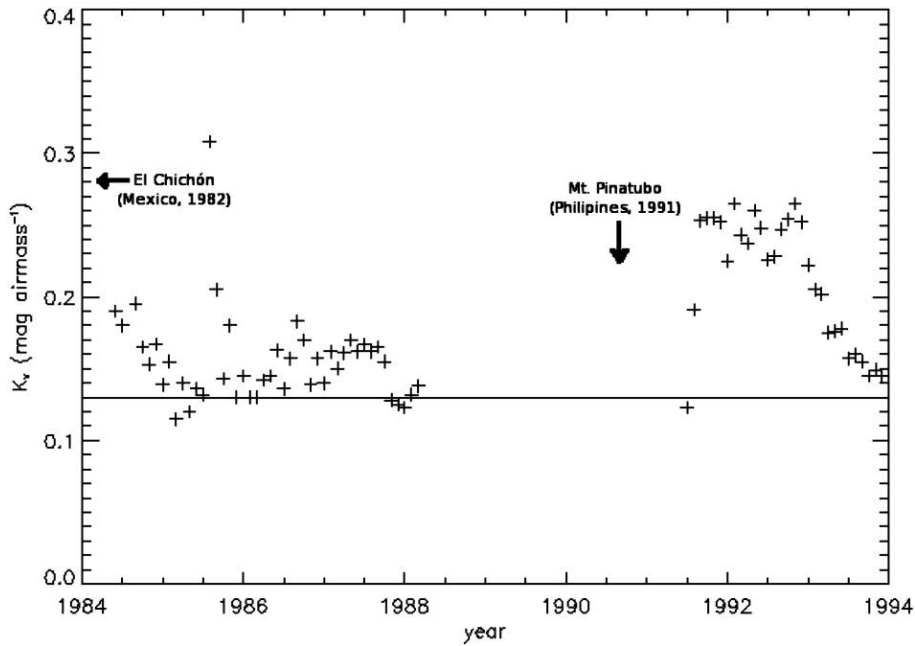

FIG. 6.—Monthly median $k_V$ for the periods affected by El Chichón (Mexico), and Pinatubo (Phillipines) volcanic ashes. The horizontal line shows the median extinction during the period not affected by volcanoes.

TABLE 3
MEDIAN $k_V$ DURING THE PERIOD AFFECTED BY MOUNT PINATUBO AND EL CHICHÓN ERUPTIONS

| Year | 1 | 2 | 3 | 4 | 5 | 6 | 7 | 8 | 9 | 10 | 11 | 12 |
|---|---|---|---|---|---|---|---|---|---|---|---|---|
| 1984 | | | | | 0.190(0.013) | 0.180(0.048) | 0.433(0.053) | 0.195(0.033) | 0.165(0.019) | 0.153(0.016) | 0.167(0.006) | 0.139(0.007) |
| 1985 | 0.155(0.021) | 0.115(0.006) | 0.140(0.063) | 0.120(0.019) | 0.136(0.020) | 0.132(0.040) | 0.308(0.045) | 0.205(0.059) | 0.143(0.017) | 0.180(0.024) | 0.130(0.016) | |
| 1986 | 0.130(0.009) | 0.130(0.005) | 0.142(0.009) | 0.145(0.007) | 0.163(0.031) | 0.136(0.014) | 0.157(0.053) | 0.183(0.043) | 0.170(0.063) | 0.139(0.006) | 0.157(0.011) | 0.140(0.009) |
| 1987 | 0.162(0.020) | 0.150(0.005) | 0.161(0.011) | 0.170(0.013) | 0.162(0.014) | 0.167(0.015) | 0.162(0.038) | 0.165(0.058) | 0.155(0.030) | 0.128(0.017) | 0.125(0.013) | 0.123(0.09) |
| 1988 | 0.132(0.007) | 0.138(0.009) | | | | | | | | | | |
| 1991 | | | | | | 0.123(0.020) | 0.191(0.029) | 0.253(0.018) | 0.255(0.018) | 0.255(0.016) | 0.252(0.020) | 0.225(0.019) |
| 1992 | 0.265(0.018) | 0.243(0.013) | 0.237(0.011) | 0.260(0.027) | 0.248(0.010) | 0.226(0.006) | 0.228(0.047) | 0.247(0.023) | 0.254(0.034) | 0.265(0.020) | 0.252(0.014) | 0.222(0.011) |
| 1993 | 0.205(0.017) | 0.202(0.024) | 0.175(0.017) | 0.176(0.014) | 0.178(0.007) | 0.157(0.015) | 0.160(0.039) | 0.155(0.036) | 0.145(0.008) | 0.149(0.027) | 0.145(0.016) | 0.150(0.026) |

NOTE.—Months shown as 1, January; 2, February, etc. Typical deviation of the median given in parentheses.

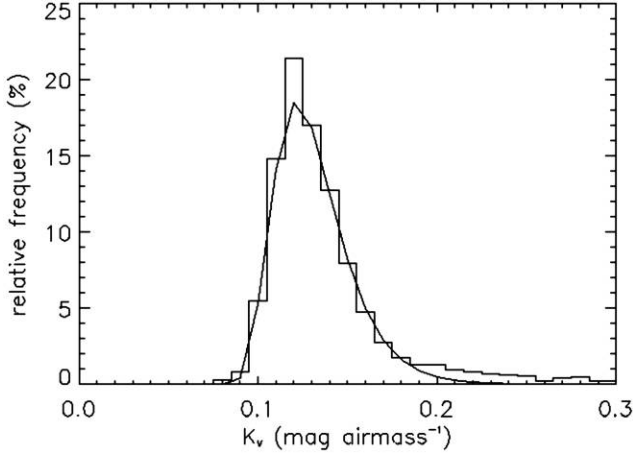

FIG. 7.—Histogram of the relative frequency of $k_V$ for nonsummer nights (October–May) compared with the best log-normal fit. Statistics compiled in Table 4.

To determine the unknown values of the log-normal fit ($\mu$ and $\sigma$), we have used the observed mean and sample variance, by taking into account the first two moments of log-normal distributions defined as

$$\text{Mean}(x - x_0) = e^{\mu + \sigma^2/2}, \tag{6}$$

$$\text{Variance}(x - x_0) = e^{2\mu + \sigma^2}(e^{\sigma^2} - 1). \tag{7}$$

We have used the method of moments, i.e., to equate the expected (or theoretical) value of the first two moments to the sample (or observed) moments, resulting in the equations

$$\bar{x} - x_0 = e^{\mu + \sigma^2/2}, \tag{8}$$

$$s^2 = e^{2\mu + \sigma^2}(e^{\sigma^2} - 1), \tag{9}$$

where

$$\bar{x} = \frac{1}{n}\sum_{i=1}^{n} x_i \tag{10}$$

is the arithmetic mean and

$$s^2 = \frac{1}{n}\sum_{i=1}^{n}(x_i - \bar{x})^2 \tag{11}$$

is the sample variance. The solution of the parameters is

$$\tilde{\mu} = 2\log(\bar{x} - x_0) - \frac{1}{2}\log(s^2 + (\bar{x} - x_0)^2) \tag{12}$$

$$\tilde{\sigma} = \sqrt{\log(s^2 + (\bar{x} - x_0)^2) - \log(\bar{x} - x_0)^2} \tag{13}$$

We have made a similar analysis using the summer nights. We have taken as $x_0$ the minimum $k_V$ observed (0.08 mag airmass$^{-1}$) minus the size of the bin (0.01 mag airmass$^{-1}$), i.e. $x_0 = 0.07$ mag airmass$^{-1}$. By testing different $k_{V(\max)}$, we have obtained a good fit to a log-normal distribution until $k_V \approx 0.155$, but there are more nights observed than predicted with $k_V > 0.155$. To obtain the best fit, we have also considered nights with slightly higher extinction ($k_V \leq 0.17$ mag airmass$^{-1}$). The extinction to which the fit begins to fail (about 0.155 mag airmass$^{-1}$) is very similar to $k^{\text{dust}}$ (the limit found by Jiménez et al. 1998 to differentiate between clear and dusty nights, which in the $V$ band is $k^{\text{dust}} = 0.153$ mag airmass$^{-1}$).

In Fig. 8 an histogram of summer nights compared with the best log–normal fit is shown. A very good agreement is observed for $k_V \leq 0.155$, as expected.

The statistics of the analysis are given in Table 4. We have made a chi-square goodness-of-fit test for the fit to a log-normal distribution of the relative frequency for nights observed in summer, and for the rest of the year separately. We have used the fact that the statistics

$$\chi^2 = \sum_{j=1}^{k} \frac{(o_j - e_j)^2}{e_j} = \sum_{j=1}^{k} \frac{o_j^2}{e_j} - k \tag{14}$$

follows a chi-square distribution with $\nu = k - 1 - m$ degrees of freedom (Spiegel 1988), where $k$ is the number of intervals in which the sample is divided, $m$ is the number of parameters of the population necessary to calculate the expected frequencies from the sample statistics, $o_j$ is the observed frequency in the $j$th interval, and $e_j$ is the expected frequency in the $j$th interval.

We have used as variable $k_V - 0.070$ mag airmass$^{-1}$ and the statistics of the analysis are given in Table 4. In the sample that excludes the summer months, we have divided $k_V$ into 8 intervals consistent with the binning (so $\nu = 5$), obtaining $\chi^2 = 6.154$, so the chance probability obtained is $Q_{\chi^2} = 0.292$. In the sample with only summer, we have divided $k_V$ into 7 intervals consistent with the binning (so $\nu = 4$), obtaining $\chi^2 = 4.613$, so the chance probability obtained is $Q_{\chi^2} = 0.329$. In both samples, the chance probability is large enough to accept the log-normal distribution as providing a good fit to the frequencies observed, even with a significant level so conservative as 0.2.

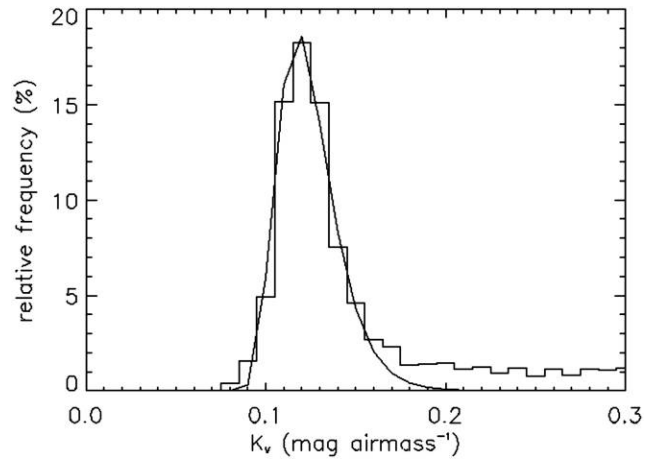

FIG. 8.—Histogram of the relative frequency of $k_V$ for summer nights (June–September) compared with the best log-normal fit. Statistics compiled in Table 4.

TABLE 4
STATISTICAL VALUES OF $k_V$ (IN mag airmass$^{-1}$) FOR DIFFERENT PERIODS

| Variable | All Data | Summer Observed | Summer Fit ($k_V \leq 0.17$) | No Summer Observed | No Summer Fit ($k_V \leq 0.205$) |
|---|---|---|---|---|---|
| Number of nights ..... | 4036 | 1685 | 1200 | 2351 | 2170 |
| Mean ................. | 0.161 | 0.183 | 0.124 | 0.144 | 0.131 |
| Median ............... | 0.130 | 0.131 | 0.123 | 0.129 | 0.127 |
| Mode ................. | 0.121 | 0.120 | 0.120 | 0.121 | 0.121 |
| Minimum ............. | 0.075 | 0.076 | 0.076 | 0.075 | 0.075 |
| Maximum ............. | 1.052 | 1.052 | 0.170 | 0.962 | 0.204 |
| Typical deviation ..... | 0.092 | 0.119 | 0.017 | 0.061 | 0.022 |
| Typ. dev./$\sqrt{n}$ ........ | 0.0014 | 0.0029 | 0.0005 | 0.0013 | 0.0005 |

If we subtract the observed summer night distribution from the fit, we obtain 495 nights out of a total of 1685 (28.8%), which can be estimated as the number of summer nights (June–September) affected by dust (or cirrus). In a similar way, the estimation of the number of nights during the rest of the year (October–May) results in 296 nights out of a total of 2351, which is 12.6% of the total and can be estimated as the number of nonsummer nights (October–May) affected by dust or cirrus. Note that over the Canary Islands, the presence of cirri is higher in winter due to the greater strength of cyclones and fronts (Erasmus & van Rooyen, 2006; González et al., 2007). Also it can be observed, mainly in winter time, that the aerosol index provided by satellites indicates the absence of aerosols whereas $k_V$ values can be larger than 0.15 (Varela et al. 2008). This situation can be explained by the presence of cirri over ORM. Photometric observations in just one filter ($k_V$) do not allow us to discriminate between clouds and dust; additional information is required.

In both the summer and nonsummer periods, the maximum of the residual extinction is set at 0.20 mag airmass$^{-1}$, which is also the modal value reported by Lombardi et al. (2008) and cited there as "the typical extinction with dust." The result is consistent with those from satellites during dust events; in these conditions, the aerosol index AI $> 0.5$–0.7 correlates with $k_V > 0.2$ (see Varela et al. 2008; Siher et al. 2004; Romero & Cuevas 2002).

## 5. USEFUL TIME

In the telescope data log there are details of observing nights lost due to technical problems or poor weather conditions for the years 1999–2003. Using this information, we have calculated the weather downtime. Some of the scheduled nights appear blank in the log and no other information is available. These nights have been discarded for the statistics.

Using the information available in the logs, we have computed average monthly weather downtime percentages. The results are given in Table 5, and in Figure 9, which also shows the mean weather downtime at the William Herschel Telescope (WHT)[3] and at ORM, from 1989–2006. A similar trend is evident. The months with the least weather downtime are May–August (less than 10%) and the months with the highest values are November, December, and January (nearly 40%). It is important to emphasize that the statistical values for the WHT are annual and constitute the most extensive database with difference (1990–2007).

A comparison with the weather downtime statistics recorded at other ORM telescopes[4] (Nordic Optical Telescope, NOT; Telescopio Nazionale Galileo, TNG; and Liverpool Telescope, LT) is shown in Table 6, which also includes details on the sampling period and operational limits.

From Table 6 it is clear that the TNG and LT record the highest values of weather downtime, and that downtime at the TNG value is the highest at ORM. The differences cannot be explained by the different sampling. The NOT and LT have a similar sampling period, and LT weather downtime is higher, and the same can be said for the TNG and the CMT. An obvious reason is the different operational criteria. The LT and TNG are more restrictive than the NOT and the WHT in their humidity and wind limits, especially the humidity, which is the main cause of weather downtime at the TNG Lombardi et al. (2007). This could partly explain why the LT and TNG have more weather downtime than either the NOT or the WHT.

Also important is that in Lombardi et al. (2007), the criteria to define a night, and a clear definition of the number of hours considered, are lacking. This operational value depends on the

---

[3] At http://www.ing.iac.es/Astronomy/observing/conditions/.
[4] At http://www.iac.es/site-testing/index.php?option=com_content&task=view&id=98&Itemid=104.

TABLE 5
AVERAGE VALUES FOR EACH MONTH (USING DATA FOR 1999–2003)

| Month | ndata | nd (weather) | nd (technical) | nd (blanks) | Weather Downtime (%) |
|---|---|---|---|---|---|
| 1 | 124 | 42 | 12 | 2 | 38.18 |
| 2 | 113 | 17 | 11 | 1 | 16.83 |
| 3 | 130 | 33 | 3 | 1 | 26.19 |
| 4 | 150 | 41 | 1 | 1 | 27.70 |
| 5 | 155 | 13 | 6 | 1 | 8.78 |
| 6 | 150 | 2 | 26 | 0 | 1.61 |
| 7 | 155 | 5 | 11 | 1 | 3.50 |
| 8 | 155 | 9 | 34 | 0 | 7.44 |
| 9 | 150 | 22 | 18 | 1 | 16.79 |
| 10 | 155 | 24 | 48 | 0 | 22.43 |
| 11 | 150 | 49 | 7 | 7 | 36.03 |
| 12 | 124 | 49 | 11 | 0 | 43.36 |

NOTE.—Number of data available in the log: ndata; number of data with poor weather: nd(weather); number of data with technical problems: nd(technical); blanks: nd(blank).

## 6. RESULTS AND DISCUSSION

Sky transparency is a key parameter for astronomical observations. Its value and statistics is of the greatest importance in defining the quality of an astronomical site. Since 1984, the ORM hosted a meridian circle telescope (the CMT), which provides nightly values of the atmosphere extinction in the visible range. In this work we have carried out an analysis of the entire CMT database available to public access.

In this article we have presented a study of the atmosphere extinction at the ORM, carried out after an analysis of data collected over more than 20 years.[5] It is based on photometric measurements in the $V$ band and the $r'$ band (transformed to the $V$-band extinction coefficient), each value using about 50 photometric standards per night, obtained from 1984 to present at the CMT. Since long-term extinction variations due to the effects of the volcanic eruptions of El Chichón (1982) and Mount Pinatubo (1991) are observed, the periods affected are not taken into account in this study.

The data affected by the volcanic ashes are analyzed separately. Although a deep interpretation in the framework of atmospheric studies is out of the scope of this work, the statistics of the periods affected by El Chichón and Pinatubo is obtained. The extinction values measured by the CMT show an increases to 0.07 mag airmass$^{-1}$ and 0.12 mag airmass$^{-1}$, due to El Chichón and Mount Pinatubo, respectively. The behavior in the case of Mount Pinatubo was simpler. The atmosphere extinction at ORM showed a sudden increase soon after the eruption, remained almost constant for about two years and smoothly decreased. Four years after the eruption, the volcanic ashes in the atmosphere, if existing, are not detected. The effects of El Chichón in the atmosphere extinction lasted 6 yr after the eruption. The volcano erupted in 1982, 2 yr before the database was created. Therefore the increase of the extinction measured in 1984 is a lower limit of the impact of this volcano on the atmosphere at La Palma, at the ORM altitude. The trend along these years is somehow more complicated, with local maxima overlapped on the more general decreasing trend. Moreover, there are two absolute maxima that coincide with two episodes of Saharan dust intrusion in 1985 August and 1986 August.

telescope itself, and the number of clear hours that can be used for operation depends on the particular telescope. In summary, we do not know how many lost hours (continuous periods or otherwise) define a night as useful or lost. This is a caveat that affect to all telescopes in general and should be considered as an intrinsic (unknown) error.

In this study we have considered only weather downtime where there were no recorded observations during a whole night. Considering the entire 5 yr period under study (56 months, from 1999 April–2003 November), we estimate the weather downtime at CMT to be 20.7%. The weather downtime is calculated as the fraction of the time lost due too poor weather conditions in the otherwise available nights.

The possible influence of the filter change (from $V$ to $r'$) and observing uncertainty change ($\pm 0.005$ to $\pm 0.0005$ mag airmass$^{-1}$) in the measurements is analyzed, although the prescribed conversion factor and an adequate binning and statistics for grouped data have been applied. We have observed that the cumulative frequency of $k_V$, using only the extinction obtained with a $V$ filter are very similar to that obtained using only extinction obtained with an $r'$ filter. The result is very similar in both cases and confirms that there is no need to apply a further correction.

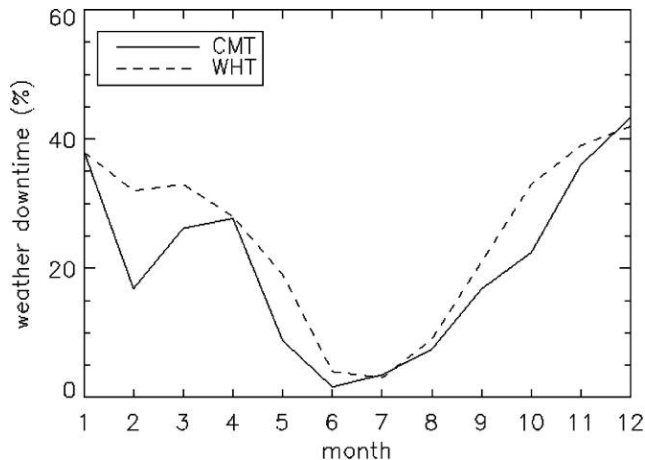

FIG. 9.—Relative monthly frequency of weather downtime at the CMT from the log for 5 yr compared with that of the WHT for 18 yr.

[5] At http://www.ast.cam.ac.uk/~dwe/SRF/camc.html.

TABLE 6
Statistical Values (as %) of Weather Downtime

| Variable | CMT (this work) | WHT | NOT | LT | TNG |
|---|---|---|---|---|---|
| Mean | 20.7 | 26.33 | 26.11 | 29.94 | 30.24 |
| Standard deviation | 19.8 | 5.73 | 16.44 | 18.74 | 20.70 |
| Maximum | 69 (1999 Nov) | 36 (2001) | 53.4 (2008 Feb) | 74.35 (2006 Jan) | 90.8 (2005 Feb) |
| Minimum | 0 (various) | 15 (2000) | 0.9 (08/2007) | 1.79 (2006 Jun) | 1.5 (2003 Jul) |
| Sampling period | 1999 Apr–2003 Nov | 1990–2007 | 2006 Oct–2008 Nov | 2006 Jan–2008 Oct | 2000 Jan–2005 Dec |
| Sampling duration | 56 months | 18 yr | 26 months | 34 months | 72 months |
| Humidity limit | Observer dependent | 90% | 90% | 80% | 85% |
| Wind limit | Observer dependent | 80 km hr$^{-1}$ | 72 km hr$^{-1}$ | 60 km hr$^{-1}$ | 54 km hr$^{-1}$ |

Considering the 20 yr span of our database (from 1988–2009, excluding 1992 and 1993), we obtain a median $k_V$ of 0.130, a mode of 0.121 and a mean of 0.161 mag airmass$^{-1}$, with an uncertainty of ±0.002 mag airmass$^{-1}$. The median extinction has not varied over the years, with a slight, not statistically significant decrease over the last 5 yr. The lack of temporal variation (during 20 yr) of the extinction contradicts those results that conclude that the number of African dust intrusions might have increased in the Canary Islands at sea level (Torres et al. 2002). An explanation to this apparent contradiction is the drainage of dust to lower altitude, falling below the level of the observations (see Varela et al. 2008 for a more complete discussion). Moreover, a recent study by Cuevas & Baldasano (2009) with updated measurements proves that there is neither an increase in the frequency of Saharan dust episodes nor an increase in their intensity at the altitude of the Canarian Observatories. It is clear that studies of the atmosphere at the ORM altitude (around 2400 m) have to be carried out with in situ instruments.

Analysis of the data has also been carried out taking into account the general synoptic situation, which divides the year into two periods, based in the seasonal trend of the trade winds. These periods are summer, which comprises June, July, August, and September, and the rest of the year.

When separating the summer and the rest of the year in the analysis, an identical median and mode $k_V$ to within errors are obtained (0.131 mag airmass$^{-1}$ in summer and 0.130 for the rest of the year for the median and 0.119 mag airmass$^{-1}$ in summer and 0.121 the rest of the year for the mode), but the mean $k_V$ is 0.183 mag airmass$^{-1}$ in summer and 0.145 during the rest of the year. This result is much better than that obtained at the 2800 m altitude at Mauna Kea in Hawaii (median $k_V$ of 0.17 mag airmass$^{-1}$ (Krisciunas 1990). However, at the 4200 m altitude at Mauna Kea, median $k_V$ is 0.11 mag airmass$^{-1}$ (Krisciunas et al. 1987), which is lower than the value at the ORM.

We have developed a method to derive the percentage of nights affected by dust or cirrus. The derived percentage is ∼29% in summer and ∼13% during the rest of the year. If we include in the analysis only dust-free and noncloudy nights, we obtain at the ORM a mean value of $k_V$ of 0.124 mag airmass$^{-1}$ in summer and 0.132 mag airmass$^{-1}$ for the rest of the year.

This value is similar (even better in summer) to that obtained at La Silla (0.130 mag airmass$^{-1}$) using MD nights (Burki et al. 1995), which are foreseeable good, stable, uninterrupted nights, assuming that these selected nights have statistical values of extinction similar to the dust-free and noncloudy nights at the ORM.

Our result points, in agreement with numerous other studies (e.g., Rodriguez et al. 2009 and references therein), to a seasonal periodicity of dust episodes, which contradicts the results presented by Lombardi et al. (2008). The reason for this could be the poor statistics of the study by Lombardi et al. (2008) together with a possible overinterpretation of the effect on atmospheric transmission of the particle distribution measured by an airborne dust monitor. Some of the particles could be of "anthropogenic origin" and not associated with dust, and with a different vertical distribution.

We have estimated the weather downtime with the available information in the data logs. Our global estimate of weather downtime for the period 1999–2003 based on the CMT log is 20.7%. This value compares reasonably well—taking into account the caveats intrinsic to the definition of weather downtime—with data provided by other telescopes at ORM, in particular with the numbers provided by the NOT and WHT. The values reported by the TNG, however, are higher, the highest of all the telescopes on the mountain, probably due to its more restrictive operational restrictions. Our result fully agrees with the report provided by the CMT team when the telescope was assigned to Observatorio de San Fernando in 2005. Their average reported value, using their information for weather downtime from 1984 to 2005, was 23% (Evans 2005).

Weather downtime is fairly variable for the different months of the year. May–August are the months with the least weather downtime (less than 10%) and the months with the highest values are November–January (almost 40%). This is consistent with the data provided by the WHT.[6] It is important to emphasize that the statistical values for the WHT are annual and are, by far, the most extensive database (1990–2007) of weather downtime.

---

[6] At http://www.ing.iac.es/Astronomy/observing/conditions/.

All the information presented in this article points to the fact that conditions at the ORM are stable with a clear seasonal trend. This makes the site an excellent testbench for exploring forecasting algorithms and implementing flexible schedule protocols.

We acknowledge Thomas Augusteijn (NOT), Chris Benn (Isaac Newton Group), David Carter (LT), and Valentina Zitelli (TNG) for providing the necessary data to make Table 6. Thanks to Claus Fabricius for his help in accessing documents and reports of the CTM. We are grateful to our anonymous referee for a careful reading of the manuscript and numerous and important comments and suggestions that have contributed to improving the paper. The referee suggested inclusion of the analysis of data affected by the volcanoes in Mexico and Phillipines, which we think makes the paper useful for a broader readership. Thanks to Terry Mahoney and Nicola Caon for English corrections.